\documentclass[journal]{IEEEtran}

\usepackage[utf8]{inputenc} 
\usepackage[T1]{fontenc}    
\usepackage{hyperref}       
\usepackage{url}            
\usepackage{booktabs}       
\usepackage{amsfonts}       
\usepackage{nicefrac}       
\usepackage{microtype}      
\usepackage{lipsum}
\usepackage{graphicx}
\usepackage{float}
\usepackage{amsmath,amssymb}
\usepackage{array}

\title{Comparison of CYGNSS and Jason-3 Wind Speed Measurements via Gaussian Processes}

\begin{document}

\author{William~Bekerman and Joseph~Guinness%
\thanks{This work was supported in part by 
the National Science Foundation, Division of Mathematical Statistics under grant numbers 1916208 and 1953088.
\textit{(Corresponding author: Joe Guinness)}.}
\thanks{The authors are with Department of Statistics and Data Science, Cornell University, Ithaca, NY 14853 USA (e-mail: guinness@cornell.edu).}
}


\maketitle

\begin{abstract}
Wind is a critical component of the Earth system and has unmistakable impacts on everyday life. The CYGNSS satellite mission improves observational coverage of ocean winds via a fleet of eight micro-satellites that use reflected GNSS signals to infer surface wind speed. We present analyses characterizing variability in wind speed measurements among the eight CYGNSS satellites and between antennas. In particular, we use a carefully constructed Gaussian process model that leverages comparisons between CYGNSS and Jason-3 during a one-year period from September 2019 to September 2020. The CYGNSS sensors exhibit a range of biases, most of them between -1.0 m/s and +0.2 m/s with respect to Jason-3, indicating that some CYGNSS sensors are biased with respect to one another and with respect to Jason-3. The biases between the starboard and port antennas within a CYGNSS satellite are smaller. Our results are consistent with, yet sharper than, a more traditional paired comparison analysis. We also explore the possibility that the bias depends on wind speed, finding some evidence that CYGNSS satellites have positive biases with respect to Jason-3 at low wind speeds. However, we argue that there are subtle issues associated with estimating wind speed-dependent biases, so additional careful statistical modeling and analysis is warranted.
\end{abstract}

\ifCLASSOPTIONpeerreview
\begin{center} \bfseries EDICS Category: 3-BBND \end{center}
\fi

\IEEEpeerreviewmaketitle



\section{Introduction}

\IEEEPARstart{W}{ind} is a crucial component of the atmosphere and climate, having 
significant implications in numerous areas of daily life, from safety and transportation to industry and science.
Recording accurate wind measurements provides critical information for precisely defining weather hazards \cite{adelekan2000survey}, building skyscrapers, and landing aircraft \cite{smith1998prediction}. Reliable wind speed observations are also consequential in allowing us to efficiently conduct crop spraying \cite{endalew2010new}, monitor the global climate \cite{eichelberger2008climate}, and avoid the obstruction of essential global shipping routes \cite{rusu2018comparative}.

Scientists and engineers have developed a diverse suite of tools for measuring wind speeds. Weather stations and buoys are often equipped with anemometers to measure wind speed directly. Many earth-observing satellites carry sensors capable of inferring wind speeds. A thorough review of satellite-based methods for measuring wind speeds is provided in \cite{young2017calibration}, which includes radiometers, scatterometers, and altimeters, which are usually attached to low-earth-orbiting satellites. In addition, geostationary satellites are capable of inferring upper-air wind speeds by detecting movements in clouds via derived motion winds algorithms \cite{derived_motion}.

The Cyclone Global Navigation Satellite System (CYGNSS) is a fleet of eight micro-satellites that use the scattered signals from existing GNSS satellites to infer wind speeds at the ocean surface  \cite{ruf2012cygnss, ruf2013cygnss}. CYGNSS is a relatively new and low-cost system that, due to its ability to distribute its sensing effort over eight satellites, has the advantage of greater spatial-temporal coverage of the oceans relative to a single-satellite system, an important feature for its mission of monitoring tropical cyclones.

Our primary goal is to study the internal variability in wind speed measurements among the eight CYGNSS satellites and across each satellite's starboard and port antennas. As noted in \cite{asharaf2021cygnss}, this variability is still under study, and further calibrations are a possibility: ``it is
more likely the differences in bias, both between antennas and
between spacecraft, are caused by residual errors in the engineering calibration, which is performed individually for each
spacecraft and antenna. This is an ongoing area of investigation by the CYGNSS project team{ . . .}'' Our secondary goal is to study differences between CYGNSS and Jason-3 wind speed measurements.

Several recent articles study statistical properties of CYGNSS wind speed measurements. CYGNSS was compared with weather model forecast winds and found to be positively biased at low wind speeds and negatively biased at high wind speeds, but no comparisons were made among individual CYGNSS satellites \cite{pascual2021improved}. In the tropics, a similar pattern of positive bias for low wind speeds and negative bias for high wind speeds, relative to hourly-averaged buoy data, was detected \cite{asharaf2021cygnss}. CYGNSS biases with respect to modeled and sensed wind speeds were assessed in the context of building a complex bias correction algorithm \cite{said2021noaa}.

The spatial-temporal patterns of low-earth-orbiting satellite observations, coupled with the inherent spatial-temporal variability in wind speeds, present a data-analytic challenge for conducting the desired comparisons in our study. 
By design, the eight CYGNSS satellites do not measure winds concurrently at the same locations, so it is difficult to draw direct comparisons between observations from most pairs of these satellites. 
Thus, we rely on repeated approximate crossings between CYGNSS and Jason-3 for indirect comparisons. Jason-3 uses reflected signals from its radar altimeter to infer wind speed and other ocean surface parameters. We believe that Jason-3 is a suitable comparison because, like CYGNSS, it retrieves a snapshot of surface wind speed, as opposed to an hourly average or a model forecast, both of which can be overly smooth. 
To maximize statistical power, it is important to make judicious use of the observations arising from the limited number of nearby CYGNSS-Jason-3 passes, while avoiding a subjective determination of which passes constitute a close-enough match. 

To address these issues, we analyze the data using carefully constructed Gaussian process models. Gaussian process models have become an indispensable tool for analyzing and interpolating scattered remote sensing data. Recent examples include the analysis of Argo float data \cite{kuusela2018locally}, Orbiting Carbon Observatory-2 data \cite{susiluoto2020efficient,katzfuss2020vecchia}, surface temperatures \cite{rayner2020eustace}, Microwave Atmospheric Satellite data \cite{8127875}, and Jason-3 wind speeds \cite{guinness2018permutation}.

Our models for CYGNSS and Jason-3 wind speeds contain bias and variance parameters that are directly related to our study goals. 
In particular, each model has parameters that are interpreted as the expected difference between CYGNSS starboard- and port-measurements and Jason-3 measurements if they had measured wind speed at the exact same location and time. 
These parameters are estimated via maximum likelihood and generalized least squares, which uses a variance-minimizing linear combination of observations to make efficient use of the available data. 
Computational challenges often associated with Gaussian process models are overcome by downsampling across time and using a state-of-the-art Gaussian process approximation implemented in the publicly available GpGp R package \cite{guinness2018gpgp}. 
The Gaussian process model and associated computational techniques are the main methodological novelties of this work. 

We find that there are significant and persistent differences among some pairs of the CYGNSS sensors of a magnitude up to 1.11 m/s. There are smaller differences between the starboard and port sensors from the same satellite. Five of the eight CYGNSS satellites have a negative bias with respect to Jason-3 measurements. No substantial differences in variances among the eight satellites were detected. These results are successfully validated against a traditional empirical analysis, which shows similar trends but higher uncertainty. In addition, we investigate the possibility that bias depends on wind speed, finding some evidence that CYGNSS measurements are larger than Jason-3 at low wind speeds, though we argue that more careful analysis is needed. We conclude the paper with a discussion of the results and suggestions for how to modify the models to study variation in the biases. All of the code necessary for reproducing our results is available in a GitHub repository at \url{https://github.com/WillBekerman/satellite-wind-speeds}.

\section{Datasets and Data Processing}

We compile one year of measurements recorded by CYGNSS and Jason-3 between September 28, 2019 and September 25, 2020, 
specifically, CYGNSS Level 2 Science Data Record Version 3.0 and Jason-3 Level-2 X-GDR Data. 
Due to missing records in the Jason-3 data during the weeks of February 1, 2020 to February 14, 2020 and June 13, 2020 to June 19, 2020, we omit these periods from our data collection, yielding 49 total weeks of satellite measurements for our analysis. 

After acquiring the data in NetCDF format, we process the data in R, retaining information about spatial location, time of measurement, wind speed, CYGNSS satellite number, and CYGNSS sensor (port vs.\ starboard). We omit any observations with missing data and standardize the times to seconds since 2020-01-01 00:00 UTC. Since CYGNSS does not record any measurements over land, we keep Jason-3 observations only over open oceans and semi-enclosed seas. The data are saved in standard R Data format. 

\begin{figure}
  \centering
  \includegraphics[scale=0.48]{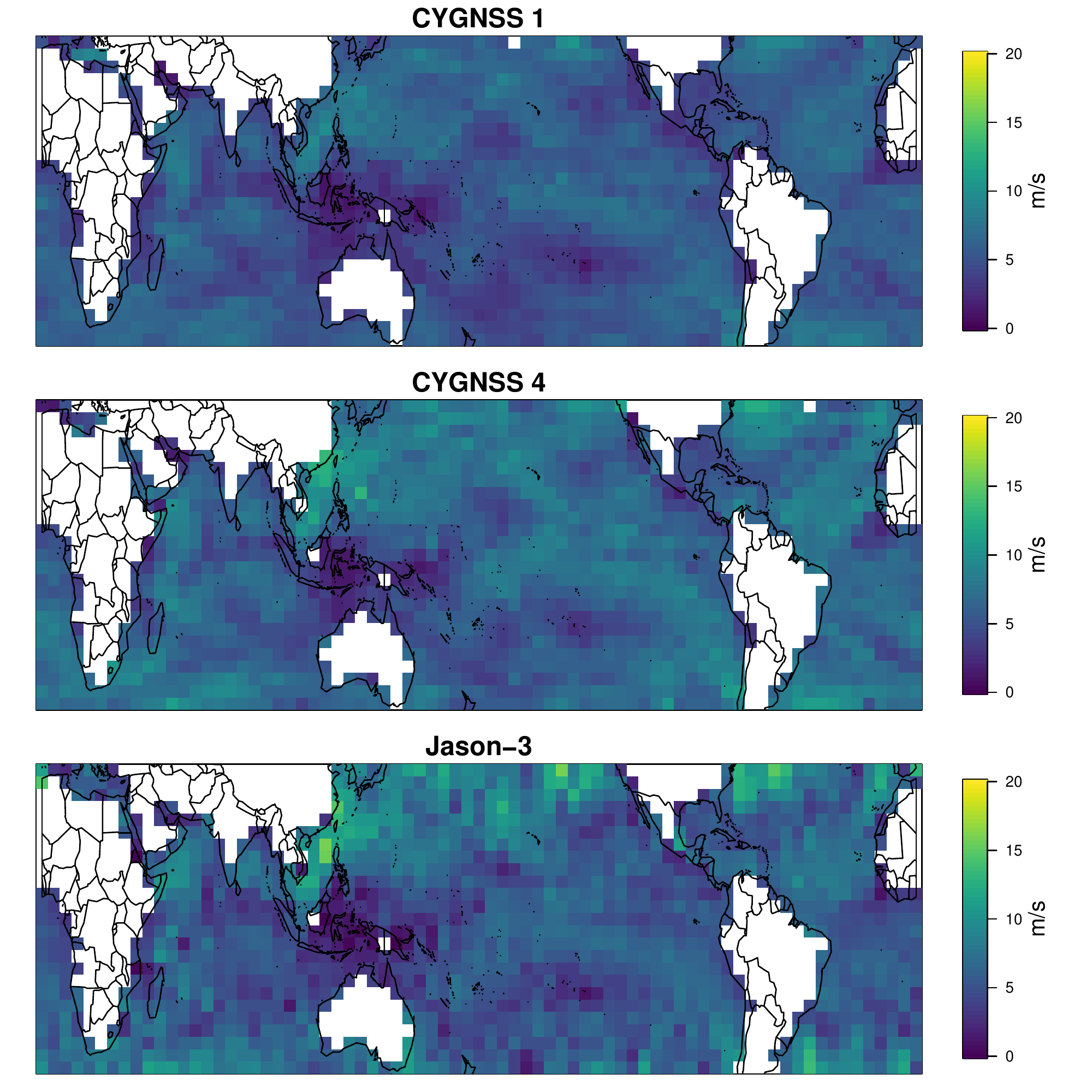}
  \caption{Wind speed (m/s) measurements recorded by CYGNSS 1, CYGNSS 4, and Jason-3 between November 30, 2019 and December 6, 2019. The value in each pixel is the sample average of all measurements falling within the pixel during the week. }
  \label{fig:cygnss-v-jason-eda-plt}
\end{figure}

In Figure \ref{fig:cygnss-v-jason-eda-plt}, we compare the wind speed measurements taken over the same latitudes by CYGNSS 1, CYGNSS 4, and Jason-3 during the week of November 30 to December 6, 2019. The value in each pixel is the average of all measurements taken within the pixel over the week. While the spatial patterns of wind speeds are similar among the three satellites, there are subtle differences. CYGNSS 4 appears to record larger wind speeds than CYGNSS 1, and Jason-3 tends to have less-smooth wind fields with more of the largest values.

\begin{figure*}
  \centering
  \includegraphics[width=1.0\textwidth]{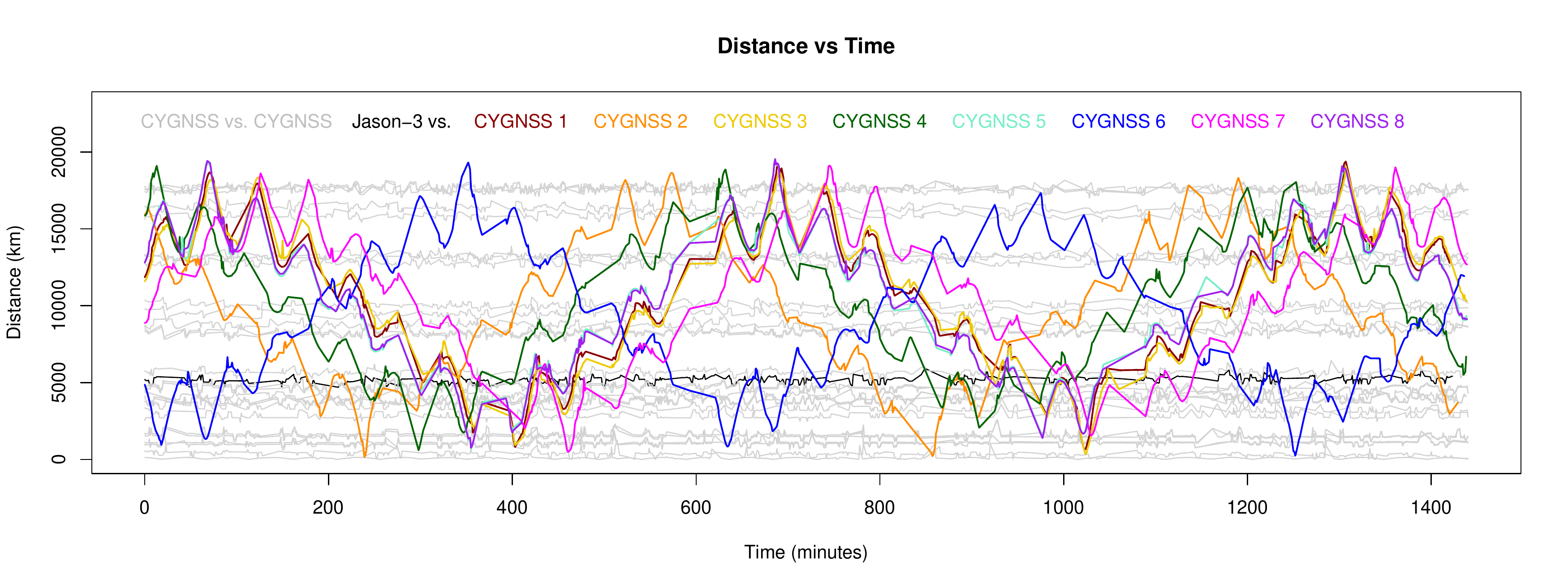}
  \caption{Distances (km) between each pair of CYGNSS micro-satellites and between Jason-3 and CYGNSS on September 29, 2019. Gray lines
  represent distances between pairs of CYGNSS micro-satellites while colored lines depict distances between Jason-3 and CYGNSS. One black line is shown to represent the distance between CYGNSS 1 and CYGNSS 4.}
  \label{fig:dist-cyg-jas}
\end{figure*}

The pixel-wise comparisons in Figure \ref{fig:cygnss-v-jason-eda-plt} can be misleading because even though the satellites have complete coverage in a one-week period, the timing of the measurements will differ among satellites, so differences could be attributable to the interaction between locally changing wind conditions and the observational times, rather than bias. 
In seeking to quantify these differences between satellite measurements, the simplest approach is to directly compare measurements recorded by CYGNSS and Jason-3 taken within small space-time windows. To explore the feasibility of the space-time matching analysis, we plot the time-varying distances between the eight CYGNSS satellites and Jason-3 over the course of one full day in Figure \ref{fig:dist-cyg-jas}.
All of the CYGNSS satellites come reasonably close to Jason-3 at one or more points during the day, with some variability in the number and proximity of such occurrences. By contrast,  
while some pairs of CYGNSS satellites nearly always measure winds at nearby locations, some pairs never do, like CYGNSS 1 and CYGNSS 4. In the next section, we propose a model designed to address the challenge of making efficient use of the CYGNSS and Jason-3 comparisons.

\section{Analysis}

\subsection{Model Description}

We first describe the statistical model used in our analysis in mathematical notation, and then provide interpretations for the model and its statistical parameters. Our analysis strategy is to fit separate models to datasets consisting of Jason-3 data and one CYGNSS satellite, and repeat the analysis for each week and each CYGNSS satellite, which means we will have many separate fits of the same general model formulation. This allows us to study whether biases are consistent over time. In the interest of keeping the number of symbols manageable, we do not provide notation for each individual model fit; the notation below is a model for an arbitrary CYGNSS satellite in an arbitrary week. Our results will contain a re-estimation of the parameters for each dataset. 

We label the $n$ observations from both CYGNSS and Jason-3 with $i=1,\ldots,n$. We use $Y_i$ for the wind speed associated with observation $i$ and model it as follows:
\begin{IEEEeqnarray}{lCr}
    Y_{i}&=\mu_i + a_{k(i)} + Z(x_i,t_i) + \varepsilon_i
\end{IEEEeqnarray}
\begin{IEEEeqnarray}{lCr}
    \mu_{i}&=b_0 + b_1 t_{i} + b_2(\mbox{lat})_{i} + b_3(\mbox{lat})^2_{i} + b_4(\mbox{lat})^3_{i}
\end{IEEEeqnarray}
\begin{IEEEeqnarray}{lCr}
    Z&\sim GP(0, K)
\end{IEEEeqnarray}    
\begin{IEEEeqnarray}{lCr}
    \varepsilon_1,\ldots,\varepsilon_n &\stackrel{ind}{\sim} N(0,\sigma^2).
\end{IEEEeqnarray}
The term $\mu_i$ is intended to capture broad-scale variation in wind speed over time $t_i$ and latitude, independent of satellite. The mapping $k(i)$ indicates which sensor produced observation $i$, with $k=1$ indicating Jason-3, $k=2$ indicating CYGNSS starboard, and $k=3$ indicating CYGNSS port. 
We model space-time variation in wind speeds with the Gaussian process (GP) $Z$, with inputs spatial location $x_i$ and time $t_i$, assumed to have mean zero and space-time Mat\'ern covariance function $\mbox{Cov}( Z(x_i,t_i), Z(x_j,t_j) ) = K( (x_i,t_i), (x_j,t_j) )$,
\begin{IEEEeqnarray}{lCr}
    K( (x_i,t_i), (x_j,t_j) )&=\mbox{Mat\'ern}(d_{ij}; \theta_1, \theta_2 ) 
\end{IEEEeqnarray}
\begin{IEEEeqnarray}{lCr}
    d_{ij} &= \Bigg( \frac{ \| x_i - x_j \|^2 }{\theta_3^2} + 
                \frac{ ( t_i - t_j )^2 }{\theta_4^2} \Bigg)^{1/2},
\end{IEEEeqnarray}
where $\theta_1$ controls the variance of $Z$, $\theta_2$ is the Mat\'ern smoothness parameter, $\theta_3$ controls the spatial-decay of the covariances, and $\theta_4$ the temporal decay.

Since $Z$ is a mean-zero process that depends only on space-time location, and not the satellite or random independent error, we interpret it as a wind speed anomaly, with the caveat that the anomaly is relative to a specific linear-in-time, cubic-in-latitude mean field. The mean field contains an intercept $b_0$, which means that $a_1$, $a_2$, and $a_3$ are not separately identifiable, but differences such as $a_2 - a_1$ and $a_3$ - $a_1$ are. As a consequence, without additional outside information, our analysis is not able to determine whether CYGNSS or Jason-3 is biased with respect to the true wind field, but it is capable of assessing whether CYGNSS and Jason-3 are biased with respect to one another via estimation of differences such as $a_2 - a_1$.

\begin{figure*}
  \centering
  \includegraphics[width=0.8\textwidth]{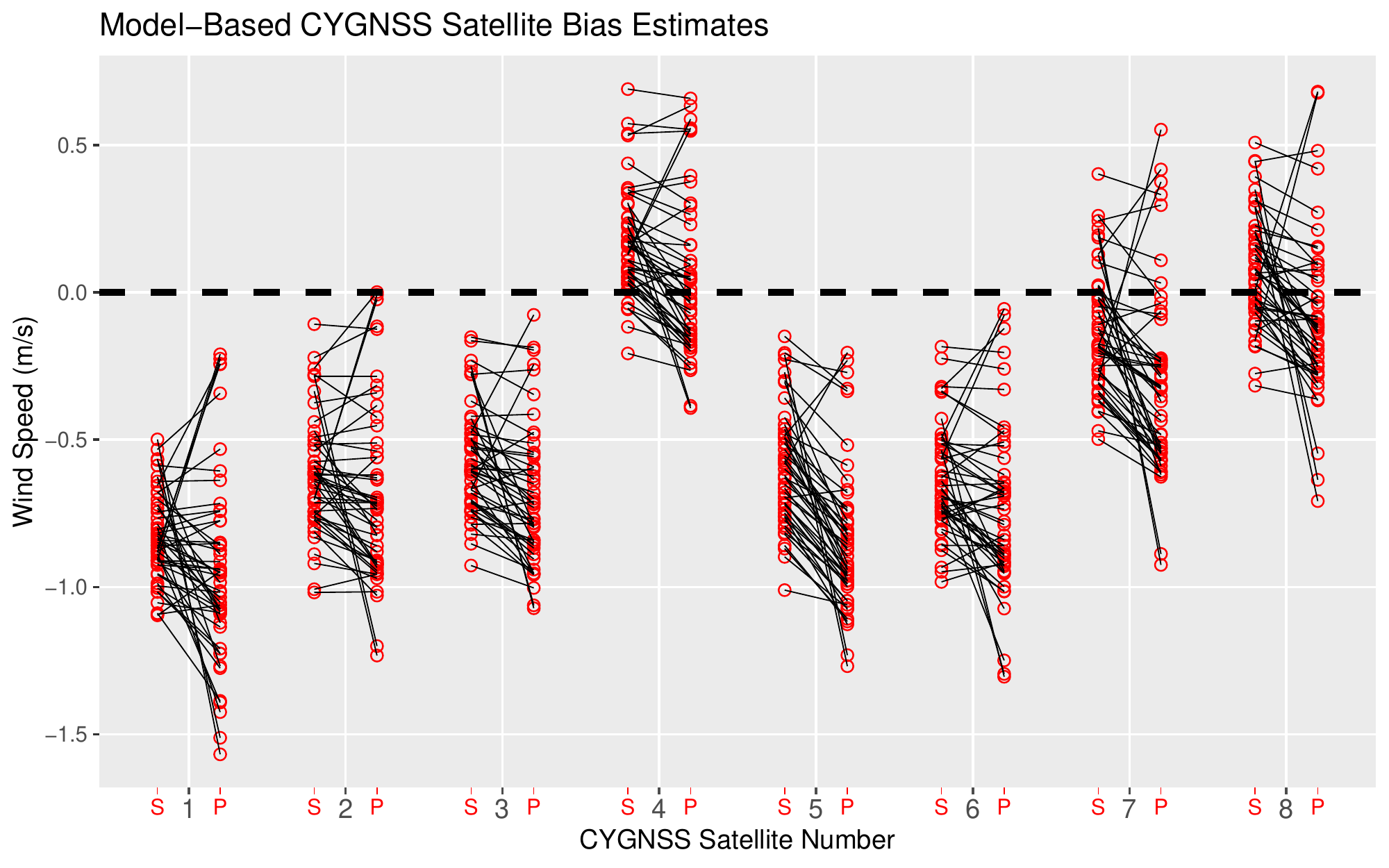}
  \caption{Model-based estimates of CYGNSS starboard (S) vs Jason-3 bias ($a_2 - a_1$) and CYGNSS port (P) vs Jason-3 bias ($a_3 - a_1$) for each CYGNSS satellite and each of the 49 weeks in our study. Starboard and port biases from the same week are connected with a black line. Discussed in more detail in Section \ref{results}.}
  \label{fig:sat-bias-plot-allweeks}
\end{figure*}

In terms of the model, we are principally interested in how CYGNSS starboard, CYGNSS port, and Jason-3 measurements would differ if they had measured wind speed at the same location at the same time. To see how this quantity relates to our model parameters, suppose that measurement $i$ was taken by CYGNSS starboard and measurement $j$ was from Jason-3, and those two measurements were recorded at the same time and location. Then
\begin{IEEEeqnarray}{lCr}\label{diff_dist}
Y_i - Y_j = a_2 - a_1 + \varepsilon_i - \varepsilon_j \sim N(a_2-a_1,2\sigma^2),
\end{IEEEeqnarray}
meaning that $a_2 - a_1$ is the bias and $2\sigma^2$ is the variance of the difference. These are the parameters of interest in our study.  By comparing the estimates of $a_2 - a_1$ and $a_3 - a_1$ across CYGNSS satellites, we can achieve our primary goal of understanding variability among the CYGNSS  measurements.

\subsection{Model Estimation}

We fit our model separately to each of the 49 weeks and each of the eight CYGNSS satellites, producing a total of 392 model fits. To reduce computational burden, we fit each model using
$20{,}000$ randomly selected observations from CYGNSS and $20{,}000$ from Jason-3, and we employ a 
popular computationally efficient Gaussian process approximation proposed by \cite{vecchia1988estimation}, implemented in the R package GpGp \cite{guinness2018gpgp}. Each model fit delivers estimates of the model parameters via maximization of the approximate likelihood function, as well as standard errors for the mean parameters. 

\subsection{Model-Based Results} \label{results}

In Figure \ref{fig:sat-bias-plot-allweeks}, for each week and each CYGNSS satellite, we plot the CYGNSS vs.\ Jason-3 bias estimates $a_{2}-a_1$ (starboard) and $a_3 - a_1$ (port). Estimates for starboard and port from the same week and CYGNSS satellite are connected by a black line. 
The bias estimates vary by week, satellite, and antenna. Five of the eight CYGNSS satellites (1, 2, 3, 5, and 6) produce negative biases with respect to Jason-3 for every week and for both antennas. The three other satellites (4, 7, and 8) have a mix of negative and positive biases across weeks. CYGNSS 1 has the largest negative biases, on average approximately -0.83 m/s for starboard and -0.94 m/s for the port antenna. Nearly all of the CYGNSS 1 biases are more negative than the most negative CYGNSS 4 bias, suggesting that CYGNSS 1 and CYGNSS 4 were persistently biased with respect to one another during our study period. Within each satellite and antenna, the bias estimates vary by roughly 0.5 to 1.0 m/s from week to week. The differences across weeks in the biases could be due to uncertainty in the parameter estimates, rather than a bias that truly varies over time. 

\begin{figure}
  \centering
  \includegraphics[scale=0.55]{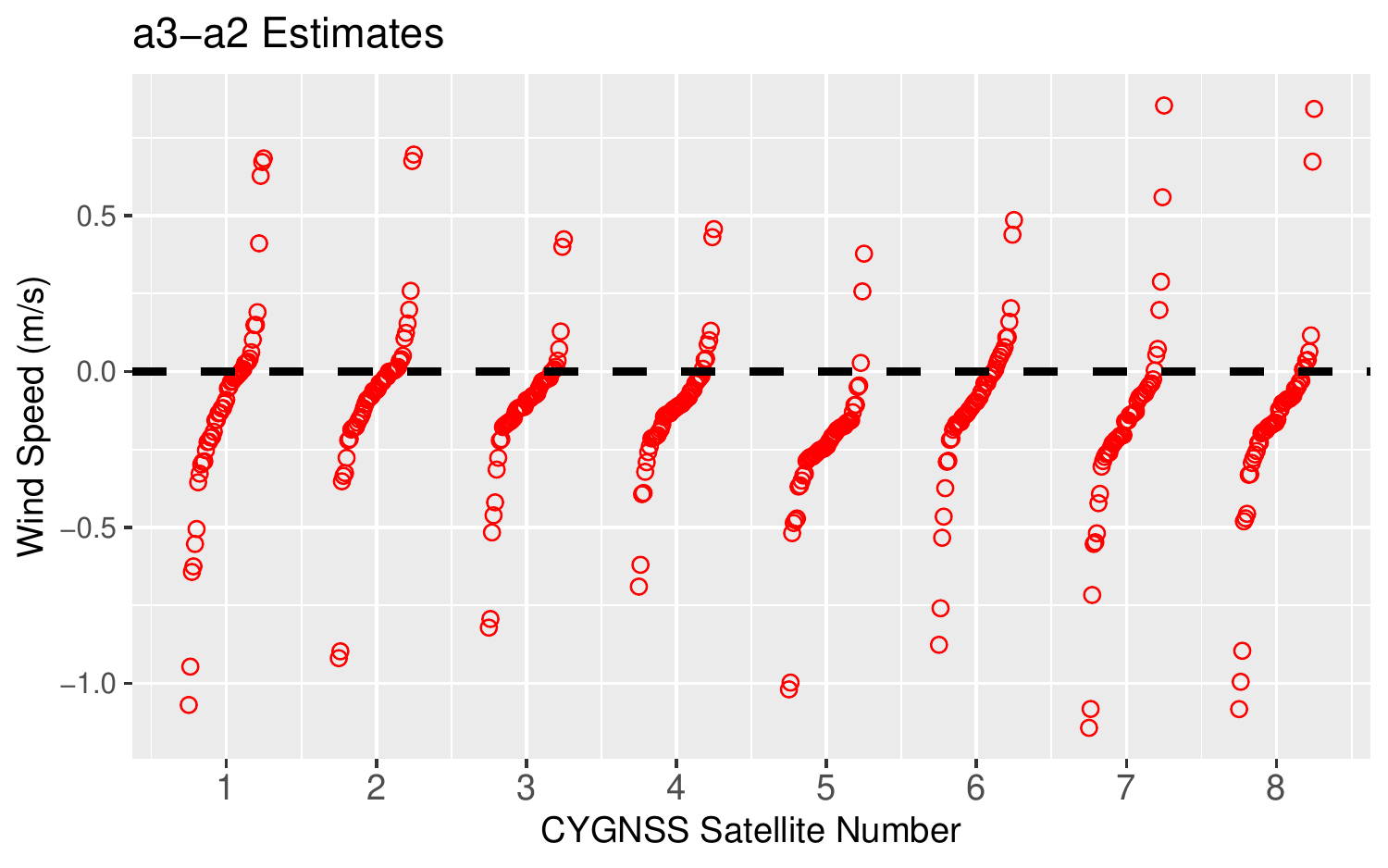}
  \caption{Estimates of $a_3 - a_2$ for each week and each CYGNSS satellite. Within each satellite, estimates are sorted and spaced horizontally to visually depict the empirical quantile function of the estimates.}
  \label{fig:sat-a3a2-plot-allweeks}
\end{figure}

By inspecting the black lines in Figure \ref{fig:sat-bias-plot-allweeks}, we observe differences between the estimates of the starboard and port biases from the same satellite within the same week. Figure  \ref{fig:sat-a3a2-plot-allweeks} explores these differences in more detail by directly plotting estimates of $a_3 - a_2$, which measure the bias between the starboard and port antennas. Most of the starboard vs.\ port bias estimates are smaller than the CYGNSS vs.\ Jason-3 biases, with most magnitudes less than 0.25 m/s, and show a mix of both negative and positive biases, though there are notably more negative than positive biases. CYGNSS 5 is the most lopsided with 46 of the 49 biases being negative.

\begin{figure}
  \centering
  \includegraphics[scale=0.55]{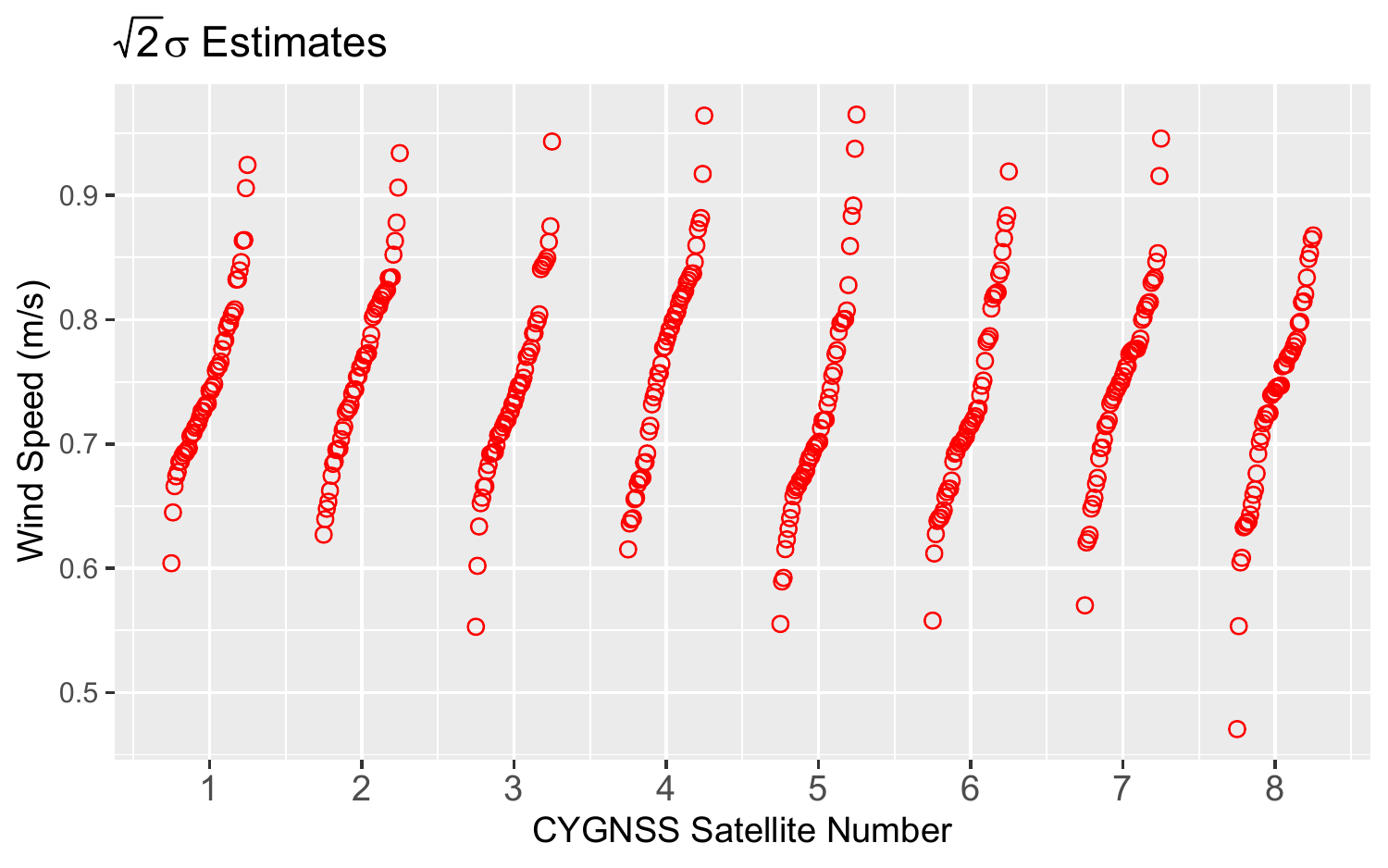}
  \caption{Estimates of $\sqrt{2} \sigma$ for each week and each CYGNSS satellite. Within each satellite, estimates are sorted and spaced horizontally to visually depict the empirical quantile function of the estimates.}
  \label{fig:sat-sigmasq-plot-allweeks}
\end{figure}

Figure \ref{fig:sat-sigmasq-plot-allweeks} displays estimates of $\sqrt{2}\sigma$, which we recall from Equation \eqref{diff_dist} is the model's standard deviation of the difference between two observations taken by different sensors at the same location and time. The estimates generally fall between 0.6 and 0.9 m/s, meaning that the size of the noise is roughly equal to the largest CYGNSS vs.\ Jason-3 biases. There is some variation of the estimates of $\sqrt{2}\sigma$ across weeks but no substantial differences among the eight CYGNSS satellites.

\subsection{Empirical Explorations} \label{empirical validation}

To validate our model-based results, we conduct additional analyses based on simple averages of differences between CYGNSS and Jason-3 wind speeds that fall within small space-time windows. For each of the eight CYGNSS satellites, each antenna, and each week between September 28, 2019 and September 25, 2020, we divide the week into two-hour windows and find the pair of CYGNSS and Jason-3 observations that are closest in distance within the two-hour window, ignoring any windows that do not have a pair that fall within 25km. We then take the average of the differences between the selected pairs of CYGNSS and Jason-3 wind speeds. We refer to the averages of these differences as our empirical bias estimates. 

\begin{figure*}
  \centering
  \includegraphics[width=0.8\textwidth]{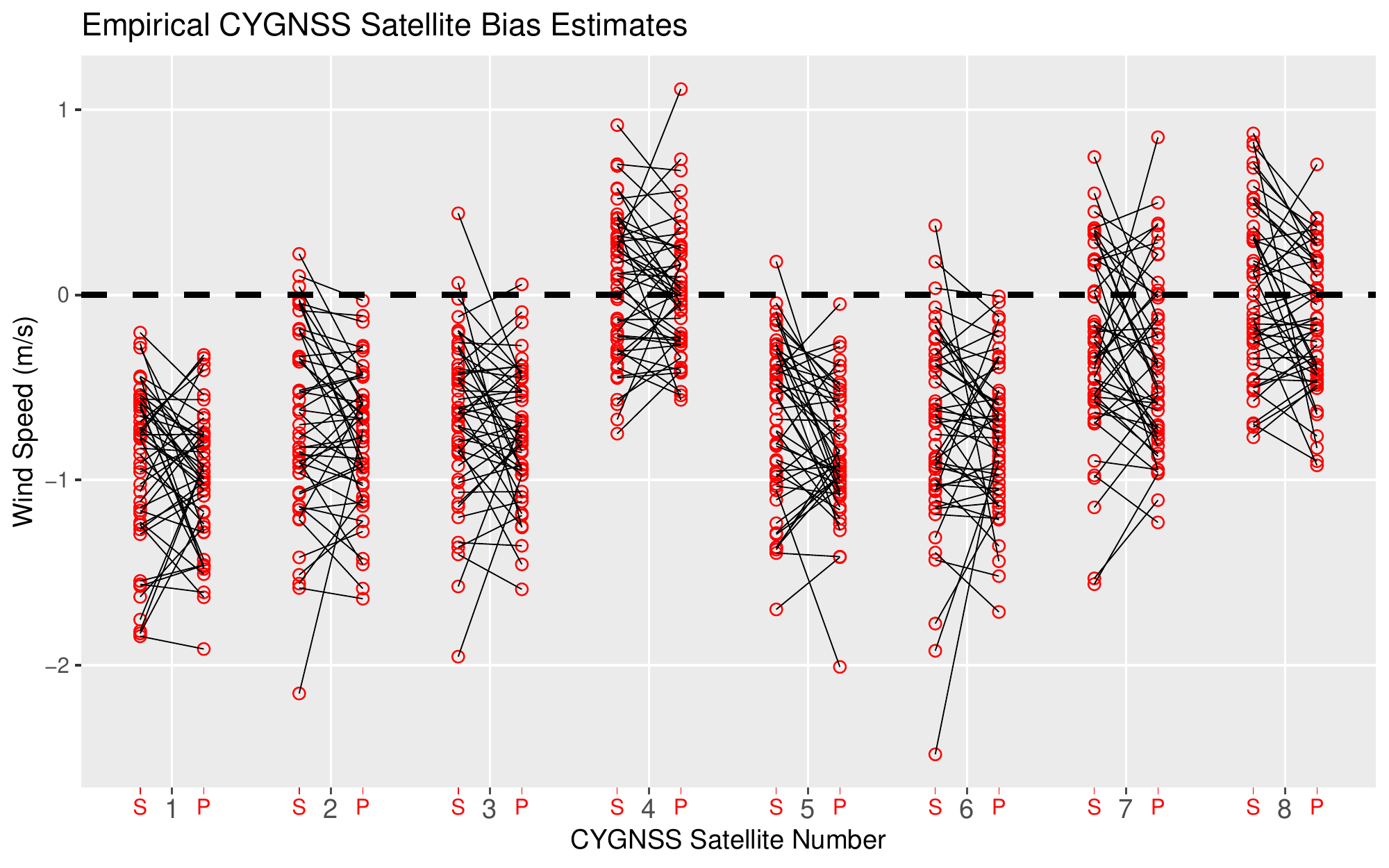}
  \caption{Empirical estimates of CYGNSS starboard (S) vs Jason-3 bias and CYGNSS port (P) vs Jason-3 bias for each CYGNSS satellite and each of the 49 weeks in our study. Starboard and port biases from the same week are connected with a black line. Discussed in more detail in Section \ref{empirical validation}.}
  \label{fig:sat-bias-plot-allweeks-empirical}
\end{figure*}

Analogously to Figure \ref{fig:sat-bias-plot-allweeks}, we plot in Figure \ref{fig:sat-bias-plot-allweeks-empirical} the empirical bias estimates over all CYGNSS satellites, antennas, and weeks. The general patterns of the empirical and model-based bias estimates are quite similar. The same five CYGNSS satellites have largely negative biases with respect to Jason-3, while the other three have a mix of negative and positive biases. The average size of the biases are similar as well, ranging roughly from -1.0 to 0.05 m/s. As in the model-based analysis, the port biases in CYGNSS 5 are, on average, more negative than the starboard biases. The empirical biases differ in that the variation across weeks is larger than in the model-based biases, which produces more overlap among the eight CYGNSS satellites and two antennas. For instance, whereas there was essentially no overlap between the CYGNSS 1 and CYGNSS 4 model-based biases, the empirical biases show more substantial overlap. In addition, the difference between the CYGNSS 5 starboard and port model-based biases is more clear than the difference between the empirical biases.

\begin{figure}
  \centering
  \includegraphics[scale=0.43]{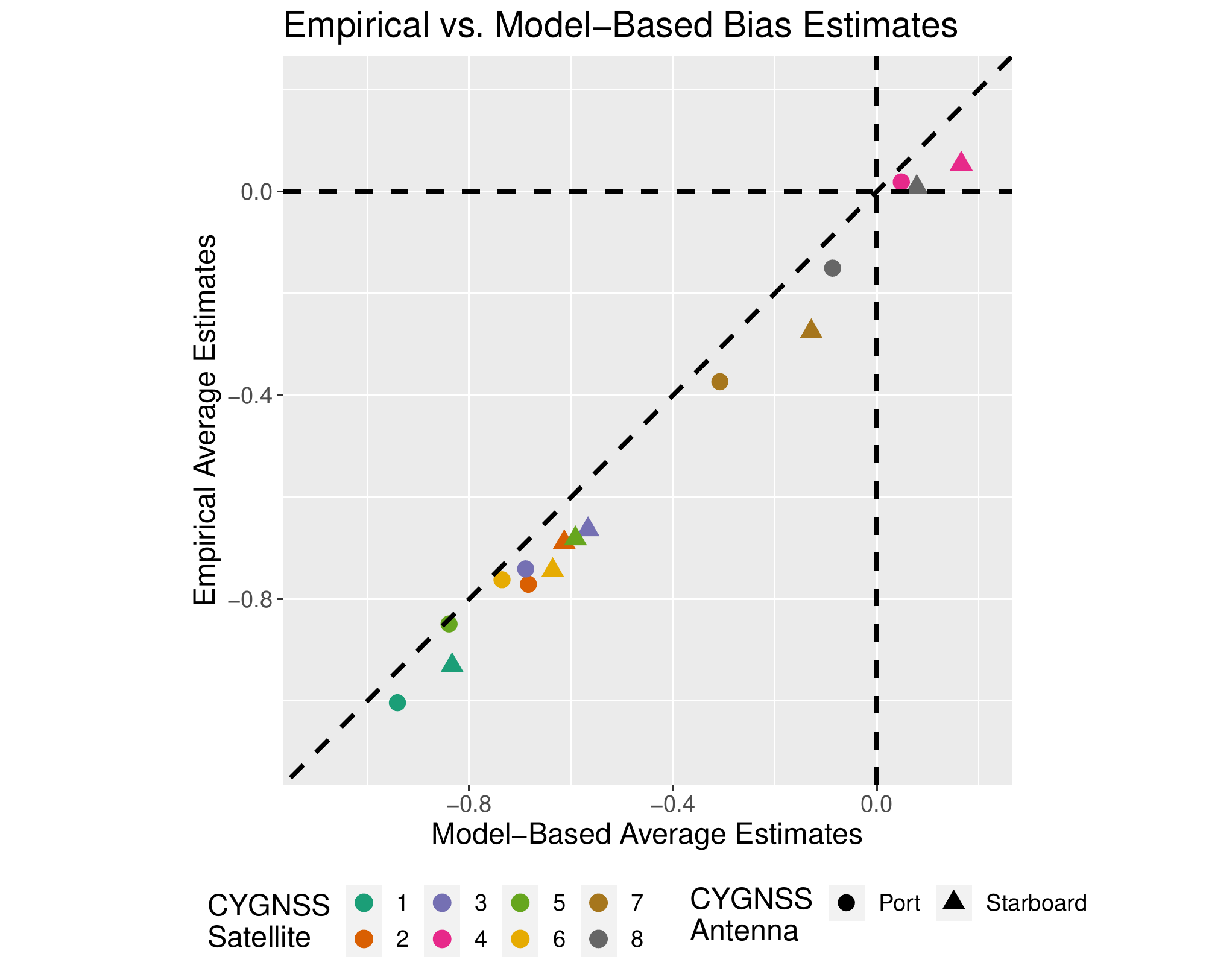}
  \caption{Average CYGNSS satellite bias estimates from our model-based approach and our empirical strategy.}
  \label{fig:empvmod}
\end{figure}

To further compare the model-based and empirical bias estimates, we average the estimates over the 49 weeks and plot them in Figure \ref{fig:empvmod}. The estimates generally follow the 45 degree line, with the model-based estimates being slightly more positive than the empirical estimates. The points tend to cluster by satellite, suggesting that the difference between starboard and port within a CYGNSS satellite is generally smaller than the differences among the eight CYGNSS satellites. Interestingly, every starboard point is northeast of its corresponding port point, indicating that the average empirical and model-based starboard biases are more positive than the corresponding port biases for every CYGNSS satellite.

\begin{figure*}
  \centering
  \includegraphics[width=1.0\textwidth]{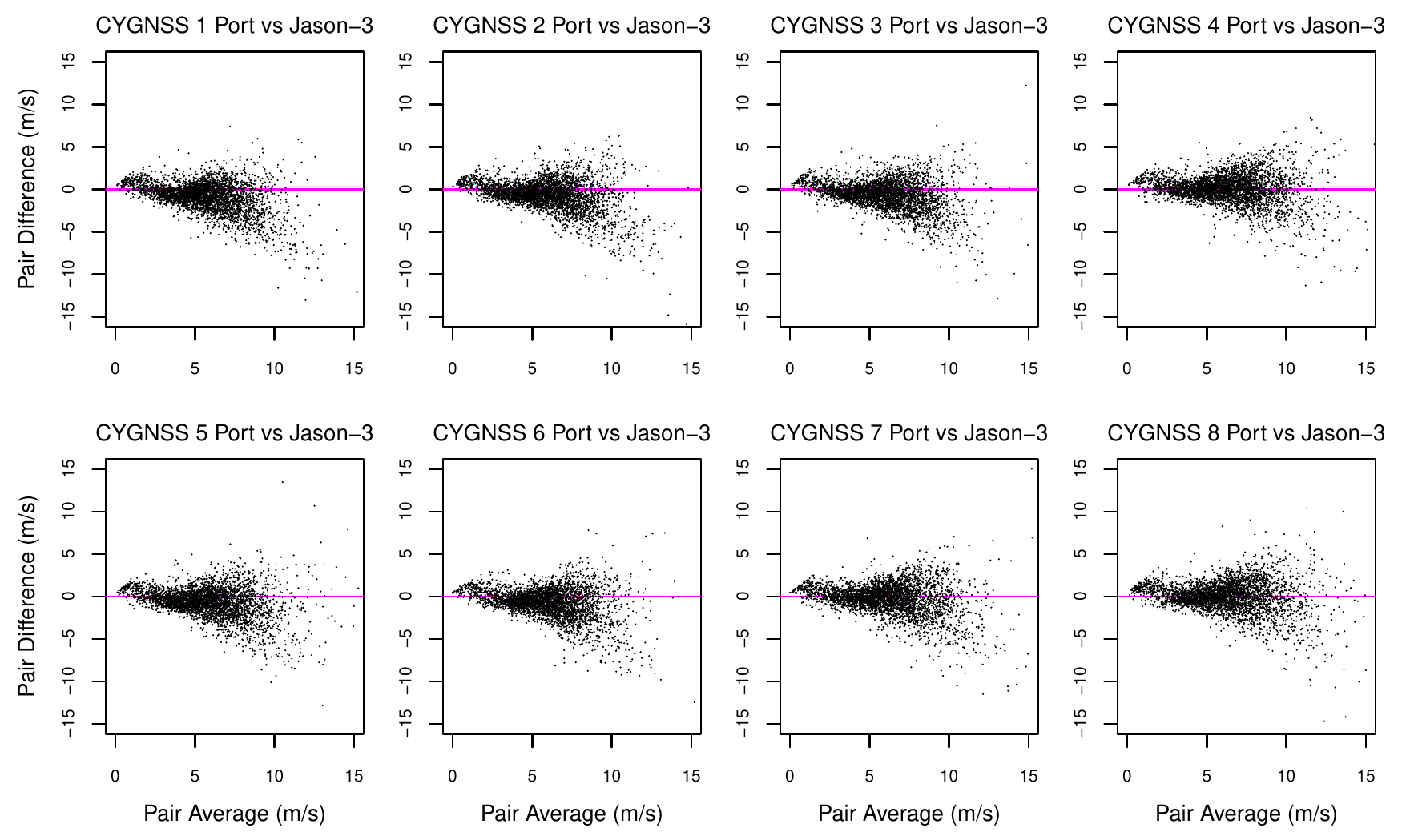}
  \caption{CYGNSS minus Jason-3 vs.\ the average of CYGNSS and Jason-3 for pairs of observations. Each pair represents the two closest observations from each two hour window over the entire year, provided that the distance is less than 25 km.}
  \label{fig:pair-average-difference}
\end{figure*}

Previous studies have explored whether bias depends on the magnitude of the wind speed. These analyses are complicated by the fact that we never have access to the ``true'' wind speed. One could take the more accurate measurement as the ``true'' wind speed and estimate bias as a function of the more accurate measurement. This approach is not without its drawbacks; when one measurement is high, the other is likely to be lower, due to standard regression-to-the-mean. To partially circumvent this issue, we plot in Figure \ref{fig:pair-average-difference} the difference between CYGNSS and Jason-3 (CYGNSS minus Jason-3) against their average. As before, we break the year into two-hour intervals and within each interval, we extract the closest pair of observations, provided that the closest distance is less than 25km. We see that among the port antennas, CYGNSS measurements are usually larger than Jason-3 for small average wind speed, but for larger average wind speeds, Jason-3 records tend to be larger. The pattern is similar for all satellites. The overall negative bias for satellites 1, 2, 3, 5, and 6 is also evident from the plots. The patterns for the starboard antennas are similar (not shown).

\section{Conclusions}

Our main finding is that during our study period of September 2019 to September 2020, persistent biases existed among the wind speed measurements recorded by the eight CYGNSS satellites and between some CYGNSS satellites and Jason-3. Considering the averages of the model-based parameter estimates over the study period, the largest bias between pairs of CYGNSS sensors was 1.11 m/s (CYGNSS 4 starboard minus CYGNSS 1 port), and the largest CYGNSS vs.\ Jason-3 bias was -0.94 m/s (CYGNSS 1 port minus Jason-3). We discovered smaller biases between the starboard and port antennas within a satellite, with the largest average bias being 0.25 m/s (CYGNSS 5 starboard - CYGNSS 5 port). 

It is not surprising to us that the two sensors on the same CYGNSS satellite would be reasonably well-calibrated with respect to one another. Since they tend to measure wind speeds at fairly close locations, direct comparison across antennas is easier. However, direct comparisons between CYGNSS satellites are more difficult due to the fact that some pairs nearly always measure wind speeds at disparate locations. Similar to other studies that make indirect comparisons with forecast winds or buoy data, we use Jason-3 as an intermediary to achieve indirect comparisons among every pair of CYGNSS satellites. We believe that Jason-3 data is appropriate because both CYGNSS and Jason-3 attempt to measure snapshots of wind speed within small space-time windows and they pass each other somewhat regularly.

These findings were facilitated by the use of a Gaussian process model that contained parameters directly related to the expected difference between measurements from different instruments taken at the same time and location. In addition to the bias parameters, the models contained a parameter related to the variance of the difference between two observations taken by different sensors at the same time and location. We did not find substantial differences in the estimates of  noise parameters among the models for the eight CYGNSS satellites. The size of the noise averaged about 0.75 m/s, meaning that the noise was of roughly equal magnitude to the largest biases, implying that about half of the mean square errors between measurements from different sensors was due to bias, half due to noise. In other words, eliminating the bias could potentially reduce the mean squared errors by a factor of two. 

We validated the model-based findings with a more traditional empirical analysis that searched for matching pairs of observations from different sensors within small space-time windows. The general pattern of the estimates in the empirical analysis was similar to that in the model-based analysis. The analyses differed in that the week-to-week variation in the empirical estimates of the bias was larger. This is expected because the model-based estimates use generalized least squares, which provides variance-minimizing parameter estimates under the assumed model.  

Some studies have suggested that the size and direction of the bias may depend on the magnitude of the true wind speed. We caution against over-interpreting these results because we do not have access to the true wind speed. If a particular sensor is chosen as the reference, we expect to see positive bias at low wind speeds and negative bias at high wind speeds due to regression-to-the-mean effects, even if the bias does not vary with wind speed. We attempted to mitigate this effect by comparing paired differences against paired averages, finding that generally the Jason-3 wind speeds increase relative to CYGNSS as the average of their two measurements increases. This aspect is certainly worthy of more exploration, with consideration of the  aforementioned statistical issues. Due to the differing accuracies of CYGNSS and Jason-3, the average may not be the best estimate of the wind speed. We could also seek out a third wind speed measurement to serve as the baseline. One could also pursue model-based estimates of biases that depend on true wind speed. To this end, consider the following extension of our model:
\begin{align*}
    Y_{i}&=\mu_i + a_{k(i)} + b_{k(i)} Z(x_i,t_i) + \varepsilon_i
\end{align*}
which contains a sensor-dependent slope multiplying the wind speed anomaly $Z(x_i,t_i)$. Inferences about biases that depend on wind speed could be obtained via estimation of $a_{k(i)}$ and $b_{k(i)}$. This is still a Gaussian process model, so we could use the same methodology to fit the model, though one would have to be careful about non-identifiability of parameters; for example, the variance of $Z(x_i,t_i)$ is not identifiable separately from the $b_{k(i)}$ parameters.

One could imagine that the bias depends on various other factors, such as latitude, time, or GNSS satellite. This sort of variation could be handled within our model framework by adding interactions between the bias and the desired factor. To capture biases that vary in space, we could extend our model as
\begin{IEEEeqnarray}{lCr}
    Y_{i}&=\mu_i + a_{k(i)} + Z(x_i,t_i) + W_{k(i)}(x_i) \varepsilon_i
\end{IEEEeqnarray}
where $W_1, W_2$, and $W_3$ are independent spatial Gaussian processes. Then we can interpret $a_2 - a_1 + W_2(x_i) - W_1(x_i)$ to be the spatially-varying starboard bias, and $a_3 - a_1 + W_3(x_i) - W_1(x_i)$ as the spatially-varying port bias. We suspect that we would need to use more than one week of data at a time to accurately estimate a spatially-varying bias.

\section{Supplementary Materials}

We run our analysis using R 4.0.5 \cite{Rcitation} on platform x86\_64-w64-mingw32/x64 running under Windows >= 8 x64. We make frequent usage of R packages "fields" \cite{fieldscitation}, "maps" \cite{mapscitation}, and "ggplot2" \cite{ggplotcitation} for creating visualizations, and "GpGp" \cite{guinness2018gpgp} for modelling.

We maintain a Github repository at \url{https://github.com/WillBekerman/satellite-wind-speeds} which contains all data, as well as R scripts to replicate our analysis.

\section*{Acknowledgment}

The authors would like to thank David Moroni for technical support downloading CYGNSS data. 

\bibliographystyle{unsrt}  
\bibliography{references}  

\begin{thebibliography}{10}

\bibitem{adelekan2000survey}
Ibidun~O Adelekan.
\newblock A survey of rainstorms as weather hazards in southern nigeria.
\newblock {\em Environmentalist}, 20(1):33--39, 2000.

\bibitem{smith1998prediction}
M~Smith and L~Chow.
\newblock Prediction method for aerodynamic noise from aircraft landing gear.
\newblock In {\em 4th AIAA/CEAS Aeroacoustics Conference}, page 2228, 1998.

\bibitem{endalew2010new}
A~Melese Endalew, C~Debaer, N~Rutten, J~Vercammen, Mulugeta~Admasu Delele,
  Herman Ramon, BM~Nicola{\"\i}, and Pieter Verboven.
\newblock A new integrated cfd modelling approach towards air-assisted orchard
  spraying. part i. model development and effect of wind speed and direction on
  sprayer airflow.
\newblock {\em Computers and Electronics in Agriculture}, 71(2):128--136, 2010.

\bibitem{eichelberger2008climate}
Scott Eichelberger, James McCaa, Bart Nijssen, and Andrew Wood.
\newblock Climate change effects on wind speed.
\newblock {\em North American Windpower}, 7:68--72, 2008.

\bibitem{rusu2018comparative}
Liliana Rusu, Alina~Beatrice Raileanu, and Florin Onea.
\newblock A comparative analysis of the wind and wave climate in the black sea
  along the shipping routes.
\newblock {\em Water}, 10(7):924, 2018.

\bibitem{young2017calibration}
IR~Young, E~Sanina, and AV~Babanin.
\newblock Calibration and cross validation of a global wind and wave database
  of altimeter, radiometer, and scatterometer measurements.
\newblock {\em Journal of Atmospheric and Oceanic Technology},
  34(6):1285--1306, 2017.

\bibitem{derived_motion}
Jaime Daniels, Wayne Bresky, Steve Wanzong, Chris Velden, and Howard Berger.
\newblock {GOES}-{R} advanced baseline imager ({ABI}) algorithm theoretical
  basis document for derived motion winds, February 2019.

\bibitem{ruf2012cygnss}
Christopher~S Ruf, Scott Gleason, Zorana Jelenak, Stephen Katzberg, Aaron
  Ridley, Randall Rose, John Scherrer, and Valery Zavorotny.
\newblock The {CYGNSS} nanosatellite constellation hurricane mission.
\newblock In {\em 2012 IEEE International Geoscience and Remote Sensing
  Symposium}, pages 214--216. IEEE, 2012.

\bibitem{ruf2013cygnss}
C~Ruf, M~Unwin, J~Dickinson, R~Rose, D~Rose, M~Vincent, and A~Lyons.
\newblock {CYGNSS}: Enabling the future of hurricane prediction [remote sensing
  satellites].
\newblock {\em IEEE Geoscience and Remote Sensing Magazine}, 1(2):52--67, 2013.

\bibitem{asharaf2021cygnss}
Shakeel Asharaf, Duane~E Waliser, Derek~J Posselt, Christopher~S Ruf, Chidong
  Zhang, and Agie~W Putra.
\newblock {CYGNSS} ocean surface wind validation in the tropics.
\newblock {\em Journal of Atmospheric and Oceanic Technology}, 38(4):711--724,
  2021.

\bibitem{pascual2021improved}
Daniel Pascual, Maria~Paola Clarizia, and Christopher~S Ruf.
\newblock Improved {CYGNSS} wind speed retrieval using significant wave height
  correction.
\newblock {\em Remote Sensing}, 13(21):4313, 2021.

\bibitem{said2021noaa}
Faozi Sa{\"\i}d, Zorana Jelenak, Jeonghwan Park, and Paul~S Chang.
\newblock The noaa track-wise wind retrieval algorithm and product assessment
  for {CYGNSS}.
\newblock {\em IEEE Transactions on Geoscience and Remote Sensing}, 60:1--24,
  2021.

\bibitem{kuusela2018locally}
Mikael Kuusela and Michael~L Stein.
\newblock Locally stationary spatio-temporal interpolation of argo profiling
  float data.
\newblock {\em Proceedings of the Royal Society A}, 474(2220):20180400, 2018.

\bibitem{susiluoto2020efficient}
Jouni Susiluoto, Alessio Spantini, Heikki Haario, Teemu H{\"a}rk{\"o}nen, and
  Youssef Marzouk.
\newblock Efficient multi-scale {G}aussian process regression for massive
  remote sensing data with satgp v0. 1.2.
\newblock {\em Geoscientific Model Development}, 13(7):3439--3463, 2020.

\bibitem{katzfuss2020vecchia}
Matthias Katzfuss, Joseph Guinness, Wenlong Gong, and Daniel Zilber.
\newblock Vecchia approximations of {G}aussian-process predictions.
\newblock {\em Journal of Agricultural, Biological and Environmental
  Statistics}, 25(3):383--414, 2020.

\bibitem{rayner2020eustace}
Nick~A Rayner, Renate Auchmann, Janette Bessembinder, Stefan Br{\"o}nnimann,
  Yuri Brugnara, Francesco Capponi, Laura Carrea, Emma~MA Dodd, Darren Ghent,
  Elizabeth Good, et~al.
\newblock The {EUSTACE} project: delivering global, daily information on
  surface air temperature.
\newblock {\em Bulletin of the American Meteorological Society},
  101(11):E1924--E1947, 2020.

\bibitem{8127875}
Weitong Ruan, Adam~B. Milstein, William Blackwell, and Eric~L. Miller.
\newblock Multiple output {G}aussian process regression algorithm for
  multi-frequency scattered data interpolation.
\newblock In {\em 2017 IEEE International Geoscience and Remote Sensing
  Symposium (IGARSS)}, pages 3992--3995, 2017.

\bibitem{guinness2018permutation}
Joseph Guinness.
\newblock Permutation and grouping methods for sharpening {G}aussian process
  approximations.
\newblock {\em Technometrics}, 60(4):415--429, 2018.

\bibitem{guinness2018gpgp}
J~Guinness and M~Katzfuss.
\newblock {GpGp}: Fast {G}aussian process computation using {V}ecchia’s
  approximation.
\newblock {\em R package version 0.4.0}, 2018.

\bibitem{vecchia1988estimation}
Aldo~V Vecchia.
\newblock Estimation and model identification for continuous spatial processes.
\newblock {\em Journal of the Royal Statistical Society: Series B
  (Methodological)}, 50(2):297--312, 1988.

\bibitem{Rcitation}
{R Core Team}.
\newblock {\em R: A Language and Environment for Statistical Computing}.
\newblock R Foundation for Statistical Computing, Vienna, Austria, 2013.

\bibitem{fieldscitation}
{Douglas Nychka}, {Reinhard Furrer}, {John Paige}, and {Stephan Sain}.
\newblock fields: Tools for spatial data, 2017.
\newblock R package version 11.6.

\bibitem{mapscitation}
Original~S code~by Richard A.~Becker, Allan R. Wilks.~R version by Ray
  Brownrigg. Enhancements~by Thomas P~Minka, and Alex Deckmyn.
\newblock {\em maps: Draw Geographical Maps}, 2018.
\newblock R package version 3.3.0.

\bibitem{ggplotcitation}
Hadley Wickham.
\newblock {\em ggplot2: Elegant Graphics for Data Analysis}.
\newblock Springer-Verlag New York, 2016.

\end{thebibliography}

\begin{IEEEbiography}[{\includegraphics[width=1in,height=1.25in,clip,keepaspectratio]{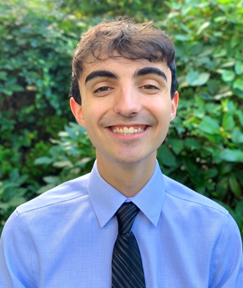}}]{William Bekerman} is currently pursuing the B.S. degree in biometry and statistics from Cornell University, Ithaca, NY, USA.

He is involved in research with the National Biomedical Center for Advanced ESR Technology and the Department of Statistics and Data Science at Cornell University, as well as the Department of Biostatistics at Columbia University. In 2019, he was involved in summer research with the Department of Applied Statistics at New York University. His research interests include statistical learning theory, high-dimensional statistical inference, spatial statistics, signal and image processing, along with applications to genomics, demography, geosciences, and other disciplines.

Mr. Bekerman is also a Student Member of the American Statistical Association and the Biophysical Society and a member of the Hunter R. Rawlings III Cornell Presidential Research Scholars program.

\end{IEEEbiography}

\begin{IEEEbiography}[{\includegraphics[width=1in,height=1.25in,clip,keepaspectratio]{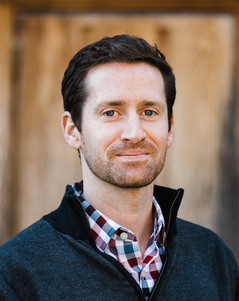}}]{Joseph Guinness} received his undergraduate degree in Math and Physics from Washington University in St.\ Louis in 2007 and his Ph.D.\ in Statistics from University of Chicago in 2012. 

He worked as a postdoctoral scholar and then Assistant Professor at North Carolina State University in the Department of Statistics from 2012 to 2017 and is now Associate Professor in the Department of Statistics and Data Science at Cornell University in Ithaca, New York.

His research interests include computational statistics for spatial-temporal data, with applications in the Earth sciences and remote sensing.

\end{IEEEbiography}

\end{document}